\documentclass[a4paper]{article}

\usepackage[numbers,sort]{natbib}

\usepackage{INTERSPEECH2020}

\usepackage{graphicx}
\usepackage{amsmath}
\usepackage{amsthm}
\usepackage{caption}
\usepackage{enumitem}
\usepackage{multirow}
\usepackage{dcolumn}
\usepackage{threeparttable}

\usepackage{hyperref}
\hypersetup{hidelinks}

\title{Listen Attentively, and Spell Once: Whole Sentence Generation via\\ a Non-Autoregressive Architecture for Low-Latency Speech Recognition}
\name{Ye Bai$^{1,2}$, Jiangyan Yi$^{1}$, Jianhua Tao$^{1,2,3}$, Zhengkun Tian$^{1,2}$, Zhengqi Wen$^1$, Shuai Zhang$^{1,2}$}
\address{$^1$NLPR, Institute of Automation, Chinese Academy of Sciences, China\\
	$^2$School of Artificial Intelligence, University of Chinese Academy of Sciences, China \\
	$^3$CAS Center for Excellence in Brain Science and Intelligence Technology}

\email{\{ye.bai, jiangyan.yi, jhtao, zhengkun.tian, zqwen, shuai.zhang\}@nlpr.ia.ac.cn}

\begin{document}

\maketitle
\begin{abstract}
Although attention based end-to-end models have achieved promising performance in speech recognition, the multi-pass forward computation in beam-search increases inference time cost, which limits their practical applications. To address this issue, we propose a non-autoregressive end-to-end speech recognition system called LASO (listen attentively, and spell once). Because of the non-autoregressive property, LASO predicts a textual token in the sequence without the dependence on other tokens. Without beam-search, the one-pass propagation much reduces inference time cost of LASO. And because the model is based on the attention based feedforward structure, the computation can be implemented in parallel efficiently. We conduct experiments on publicly available Chinese dataset AISHELL-1. LASO achieves a character error rate of $6.4\%$, which outperforms the state-of-the-art autoregressive transformer model ($6.7\%$). The average inference latency is $21$ ms, which is $1/50$ of the autoregressive transformer model.
\end{abstract}

\vspace{-5pt}
\section{Introduction}
\vspace{-5pt}

Attention based sequence-to-sequence (Seq2Seq) speech recognition systems have achieved promising performance these years \cite{bahdanau2016endtoend,kim2017joint,chan2016listen}. In these models, an encoder encodes acoustic features into high-level representations. And a decoder is a conditional language model, which predicts the next token in terms of the previous tokens and the acoustic context. At inference stage, the decoder finds the most likely token sequence approximately with beam-search algorithm. This paradigm shows powerful ability for sequence generation. However, even with non-recurrent structures (transformers) for parallelization \cite{vaswani2017attention,dong2018speech}, the autoregressive manner still affects inference speed.

Non-autoregressive Seq2Seq models were proposed for speeding up the inference of machine translation systems \cite{gu2017non,lee2018deterministic,ghazvininejad2019mask,ma2019flowseq}. These models also use an encoder-decoder architecture with attention mechanism. But they can predict all tokens in parallel rather than in step-by-step manner. It avoids multi-pass forward propagation of the decoder in beam-search, so the inference time cost is much reduced. However, the performances of these models fall behind the state-of-the-art autoregressive models. Recently, a transformer based non-autoregressive speech recognition model was proposed \cite{chen2019non}. These models use a mask-predict manner, i.e., several tokens are replaced by the mask token randomly. And during inference, the token sequence is generated by filling the masked tokens iteratively. This method uses the predicted tokens as the language context. However, it still requires multi-pass forward propagation of the decoder to complete all the masked tokens.

We believe that the language semantic\footnote{In this paper, we refer to the relationship among tokens as language semantic.} is contained in the speech signal implicitly. So, if this semantic can be extracted well, the token sequence can be generated without relying on the explicit language modeling, e.g., autoregresive language models and masked language models. In this paper, we propose a \textit{simple and effective} non-autoregressive model called LASO (Listen Attentively, and Spell Once\footnote{This name is inspired by \cite{chan2016listen}.}). We use the feedforward self attention mechanism \cite{vaswani2017attention} as basic blocks to build three modules of LASO: the encoder, the position dependent summarizer (PDS), and the decoder. The encoder encodes the acoustic features into high-level representations. The PDS summarizes the semantic at each position from the high-level representations and bridges length gap between speech and token sequence. The decoder captures token-level semantic and predicts tokens. We conduct experiments on a publicly available Chinese dataset AISHELL-1 \cite{bu2017aishell}. The proposed LASO achieves $6.4\%$ of character error rates on test set, which is better than chain model \cite{povey2016purely} and state-of-the-art autoregressive transformer models \cite{karita2019comparative}. And compared with the strong baseline autoregressive transformer model, the inference of LASO speeds up by $50\times$.

%This paper is organized as follows. \autoref{sec:bg} introduces the background of the work. \autoref{sec:laso} describes the proposed LASO model. \autoref{sec:exp} describes the experiments. At last, \autoref{sec:conc} concludes the paper and describes the future works.

\vspace{-5pt}
\section{Background}
\vspace{-5pt}
\label{sec:bg}

Speech recognition aims to convert an acoustic feature sequence to the corresponding textual token (word, sub-word, or phone) sequence. Given a speech-text pair $(x, y)$, where $x$ denotes the acoustic feature sequence, and $y$ denotes the token sequence, the autoregressive Seq2Seq model estimates the conditional probability $P(y|x)$ by decomposition with the chain rule:
\begin{equation}
\label{eq:auto}
P(y|x) = P(y_1|x)\prod_{i=2}^{L} P(y_i|y_{<i}, x),
\end{equation}
where $y_i$ denotes the token at step $i$, $y_{<i}$ denotes the subsequence $[y_1, \cdots, y_{i-1}]$, and $L$ denotes the length of token sequence. For an autoregressive model, the prediction of one token relies on the previously predicted tokens at inference stage.

The non-autoregressive Seq2Seq models assume that each token is independent from the others:
\begin{equation}
\label{eq:non-auto}
P(y|x) = \prod_{i=1}^{L} P(y_i|x).
\end{equation}
Because the prediction does not depend on other tokens, the non-autoregressive Seq2Seq model can predict the token at each step in parallel.

Token relationship is important for sequence generation. For the CTC based models \cite{graves2006connectionist}, the token relationship is usually modeled by an external language model to improve performance. For RNN-Transducers \cite{graves2012sequence}, the token relationship is modeled with a prediction network. And for attention based encoder-decoder models, the token relationship is modeled with the decoder autoregressively. The main challenge of the non-autoregressive Seq2Seq model is: can a model generate token sequence without the explicit token relationship? We believe that the token relationship is contained in the speech implicitly. If we achieve token-level representations from speech, we can generate the token sequence with the non-autoregressive model.

\begin{figure}[!t]\centering
	\includegraphics[width=0.9\columnwidth]{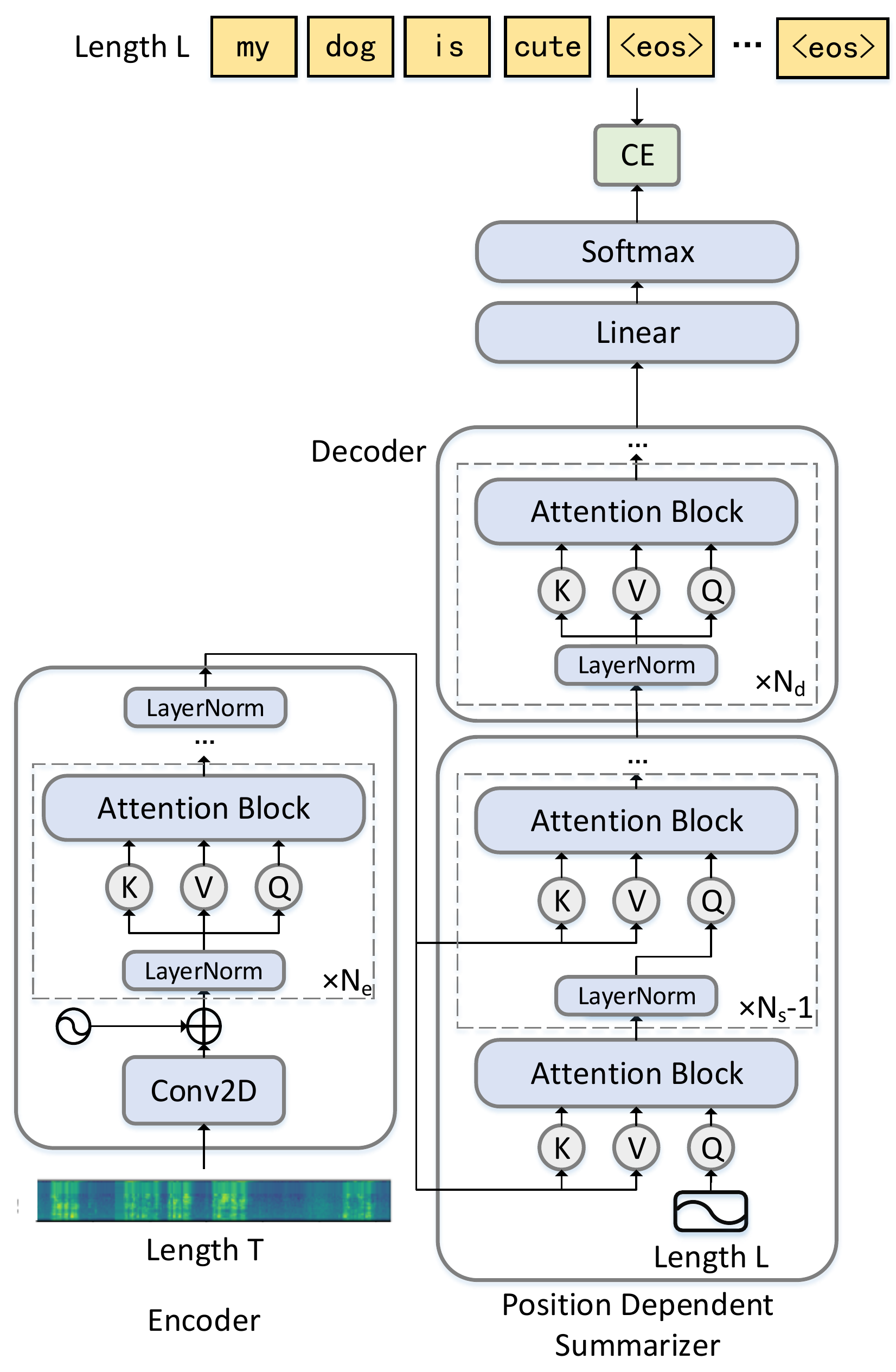} 
	\caption{The architecture of LASO. The encoder first uses a two-layer CNN to subsample the acoustic feature sequence, and then uses a stack of attention blocks to obtain high-level representations. The position dependent summarizer (PDS) queries the high-level representations for each position. It bridges the length gap between the speech sequence and the token sequence. The decoder further refines the queried outputs of the PDS. With a linear transformation, the softmax function gives the probability distribution on vocabulary at each position. The tail of the token sequence is filled by ``\texttt{<eos>}" token. During inference, LASO directly selects the most likely token, and removes ``\texttt{<eos>}". The network is trained with cross entropy.}
	\label{fig:model}
	\vspace{-20pt}
\end{figure}

%\begin{figure}[!t]\centering
%	\includegraphics[width=0.9\columnwidth]{figs/arch-nobert} 
%	\caption{The architecture of LASO.}
%	\label{fig:model}
%%	\vspace{-25pt}
%\end{figure}

\begin{figure}[!t]\centering
	\includegraphics[width=0.4\columnwidth]{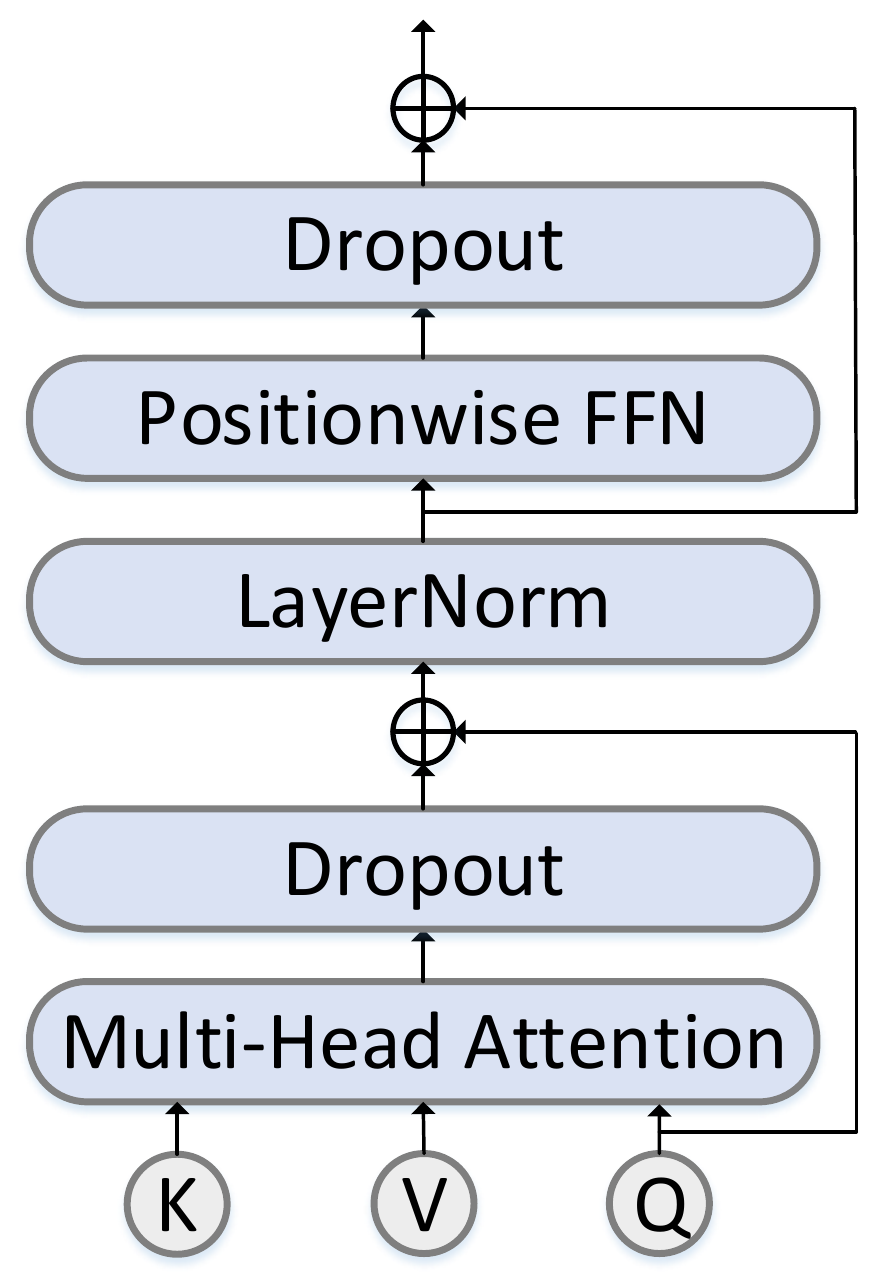} 
	\caption{An illustration of an attention block.} 
	\label{fig:block}
	\vspace{-10pt}
\end{figure}

\vspace{-5pt}
\section{The LASO Model}
\vspace{-5pt}

\label{sec:laso}
The basic idea of LASO is that the acoustic feature sequence contains not only features for pronunciation but also language semantic, i.e., relationship among tokens. If we extract representations from the whole acoustic feature sequence for each token position, we can do position-wise token prediction. Because the prediction relies on the acoustic feature sequence rather than other tokens, it can be implemented in parallel. 

Based on this idea, we formulate the position-wise token prediction as 
\begin{equation}
\label{eq:LASO}
\begin{split}
z &= \text{Encode}(x),\\
P(y_i|x) &= \text{SummarizeAndDecode}(z), i = 1, 2, \cdots, L\\
\end{split}
\end{equation}
where $z=[z_1, \cdots, z_T]$ is the hidden representation sequence which has the same length with the subsampled acoustic feature sequence $x$. To predict the token sequence with length $L$, $z$ is summarized and decoded for each position in the sequence. To generate the token sequence, the most likely token at each position is selected. The token ``\texttt{<eos>}" is added into the vocabulary as a filler for padding the token sequence to length $L$. Ideally, the tail of the generated token sequence are all ``\texttt{<eos>}'', and they are easily removed after inference. 

The proposed LASO consists of three modules. Each module consists of several attention blocks. The encoder encodes the acoustic feature sequence into high-level representations. The PDS summarizes the high-level representations to token-level representations based on the sinusoidal position encodings. The decoder generates the token for each position. The structure of the model is shown in \autoref{fig:model}. We first introduce the attention block. Then, we introduce each module of the model.

\vspace{-5pt}
\subsection{Attention Block}
\vspace{-5pt}
\label{sec:att}

Attention mechanism has been used to model global dependency in a sequence successfully \cite{raffel2015feed,vaswani2017attention}. Different from recurrent neural networks which represent context step-by-step, attention mechanism fuses all representations in a sequence by weighting sum. So, it can be computed in parallel. The dot-product self-attention is denoted as 
\begin{equation}
\label{eq:dot-product}
\text{Attention}(Q, K, V) = \text{Softmax}(\frac{QK^{T}}{\sqrt{D_k}})V,
\end{equation}
where $Q \in \mathbb{R}^{T_q \times D_{k}}$, $K \in \mathbb{R}^{T_k \times D_{k}}$, and $V \in \mathbb{R}^{T_k \times D_{v}}$ denote queries, keys, and values, respectively, $T_q$ is query sequence length, $T_k$ is key sequence length, and $D_{v}$ is the dimensionality of the keys. In this paper, $D_v$ equals to $D_k$, which is set to $D_m$ representing the model dimensionality. It can be extended to multi-head version, i.e., the hidden representations are projected into different subspaces for attention, and are concatenated together after attention \cite{vaswani2017attention}:
\begin{equation}
\label{eq:multi}
\begin{split}
&\text{MHA}(Q, K, V) = \text{Concat}(h_1,\cdots,h_H)W^o, \\
&h_i = \text{Attention}(QW_i^q, KW_i^k, VW_i^v), i = 1,\cdots,H.
\end{split}
\end{equation}
where $H$ is the number of heads, $W^o$, $W_i^q$, $W_i^k$, and $W_i^v$ are parameter matrices. The dimensionality does not change after the multi-head attention.

The position-wise feedforward network is after the attention:
\begin{equation}
\label{eq:ffn}
\text{FFN}(x) = W_2 \text{Activation} (W_1 x + b_1) + b_2,
\end{equation}
where $x$ is a vector at one position, $W_1$, $W_2$, $b_1$, and $b_2$ are learnable parameters, $\text{Activation}$ is a nonlinear activation function. In this work, we use gated linear units (GLUs) \cite{dauphin2017language}. Residual connection \cite{he2016deep} and layer normalization \cite{ba2016layer} are used in the attention block. We use pre-norm mechanism for stable training \cite{nguyen2019transformers}. The attention block is the basic component of LASO.

\vspace{-5pt}
\subsection{Encoder}
\vspace{-5pt}

The first part of the encoder consists of a two layers of convolutional neural network (CNN) for capturing locality of in the feature sequence. The stride of each CNN layer is $2$, so it also subsamples frame rates and compress the length of the sequence to $1/4$. Following \cite{vaswani2017attention}, we add sinusoidal position for self attention mechanism to capture the order. Then, $N_e$ attention blocks are used for capturing long-term relationship. Keys, queries, and values are all the inputs, i.e., self attention.

\vspace{-5pt}
\subsection{Position Dependent Summarizer}
\vspace{-5pt}

The main gap between the acoustic feature sequence and the textual token sequence is the length. Specifically, a textual token is a highly compressed semantic representation, and multiple acoustic feature frames correspond one textual token. To address this, we propose a PDS module to summarize the representations from the encoder, and to re-organize them in terms of the token positions. Basically, it is also composed of a stack of attention blocks, but the keys and the values are the outputs of the encoder. The queries of the first block are position encodings with maximum length $L$, and the queries of the follow-up blocks are the outputs of the previous block, as shown in \autoref{fig:model}. We use sinusoidal functions \cite{vaswani2017attention} to encode positions:
\begin{equation}
\label{eq:pe}
\begin{split}
\text{pe}_{i, 2j} &= \text{sin}(i / 10000^{2j/D_m}),\\
\text{pe}_{i, 2j+1} &= \text{cos}(i / 10000^{2j/D_m}), \\
\end{split}
\end{equation}
where $i = 1, \cdots, L$ denotes the $i$-th position, and $2j$ and $2j+1$ denote element indexes in a vector. The sinusoidal position encodings provide position dependent information to query representation corresponding to specific position in token sequence from the encoder outputs. So, the sequence length matches the token sequence, i.e., the length of the outputs of PDS is $L$. $L$ can be set by counting the lengths in the training set.

\vspace{-5pt}
\subsection{Decoder}
\vspace{-5pt}

After the PDS, we use the decoder to further capture token relationship. The outputs of the PDS can be seen as the representations corresponding to the tokens. So, we use self attention mechanism to capture the semantic relationship in the sequence. The decoder leverages a stack of attention blocks, and the keys, values and queries are the outputs of the previous block. After the decoder, we use a linear transformation to project the self attention based semantic representation, and softmax functions to compute probability distributions on the token vocabulary for each position.

\vspace{-10pt}
\subsection{Learning}
\vspace{-5pt}

For optimizing the parameters of the model, we minimize the position-wise cross entropy loss 
\begin{equation}
\label{eq:ce}
\text{CE}(\theta) = -\frac{1}{NL} \sum_{(x, y) \in \mathcal{D} } \sum_{i=1}^{L} \log P(y_i|x; \theta).
\end{equation}
where $\mathcal{D}$ is the dataset which contains $N$ pairs of speech and token sequence $(x, y)$, $L$ is the maximum length we pad to, and $y_i$ is the token at position $i$ in token sequence $y$.

\vspace{-10pt}
\subsection{Inference}
\vspace{-5pt}
For decoding, we just select the token which has the highest probability at each position. Given an acoustic feature sequence, the predicted token at position $i$ is 
\begin{equation}
\label{eq:dec}
\hat{y_i} = \arg\max_{y_i} P(y_i | x; \theta). \quad i=1,\cdots,L,
\end{equation}
After prediction, the filler tokens ``\texttt{<eos>}" at the tail of the sequence are removed.

\begin{table}[!t]	
	\caption{The description of AISHELL-1. ``Length'' means the average number of tokens per utterance.}
	\centering
	\begin{tabular}{ | l || c | c | c | c | } 
		\hline
					 & \#Utter.  & \#Hour    & Length & \#Speaker       \\
		\hline
		\hline
		Training     & $120,098$ & $150$     & $14.4$     &    $340$           \\
		\hline
		Development  & $14,326$  & $18$      & $14.3$     &    $40$            \\
		\hline
		Test         & $7,176$   & $10$      & $14.6$     &    $20$            \\
		\hline
	\end{tabular}
	\label{tab:dataset}	
	\vspace{-15pt}
\end{table}

\vspace{-5pt}
\section{Experiments}
\vspace{-5pt}

\label{sec:exp}
\vspace{-5pt}
\subsection{Datasets}
\vspace{-5pt}

We conduct experiments on a publicly available Chinese Mandarin corpus AISHELL-1 \cite{bu2017aishell}. The dataset includes about $150$ hours of speech for training, about $18$ hours of speech for development, and about $10$ hours speech for test. The speakers of training set, development set, and test set are not overlapped. All the recordings are in $16$ kHz WAV format.
\vspace{-5pt}
\subsection{Experimental Setup}
\vspace{-5pt}

We use $80$-dimension Mel-filter bank features (FBANK) as the input, which are extracted every 10ms with 25ms of frame length. The token vocabulary contains $4231$ characters in training set and two special symbols, i.e., ``\texttt{<unk>}" for unseen characters and ``\texttt{<eos>}" as the filler of the tail of a token sequence.

The structure of the LASO model is shown in \autoref{fig:model}. Each layer of the two-layer subsampling CNN consists of $32$ convolution filters with size $3\times3$, and the stride on time axis is $2$. The activation functions of the CNN are ReLUs. All the attention blocks used in the model are the same. Both the encoder and the decoder have $6$ attention blocks, i.e., $N_e=N_d=6$ in \autoref{fig:model}. All the attention blocks have $8$ heads. We compare different numbers of the attention blocks of PDS, i.e., $N_s=1,2,3$ and $4$, respectively. The intermediate dimensionality of the position-wise feedforward network is $2048$, and the activation function is GLU. We train two types of LASO with different model dimensionalities $D_m$. We refer to the model with $D_m=512$ as LASO-base, and the model with $D_m=768$ as LASO-big. 

We re-implement Speech-Transformer as the baseline \cite{dong2018speech}. It uses the same CNN as our LASO architecture. Following their configuration, both encoder and the decoder have $6$ layers. The dimensionality of the model is $512$, and the intermediate dimensionality of the position-wise feedforward network is $2048$. The number of heads of the multi-head attention is $8$.

All models are trained with the same procedure. We use the Adam algorithm to train the model for $130$ epochs. Each batch contains about $100$ seconds of speech, and we accumulate gradients of $12$ steps for simulating big batch \cite{ott2018scaling}. We follow the warm-up learning rate schedule \cite{vaswani2017attention}:
\begin{equation}
\alpha = D_{\text{m}}^{-0.5}  \cdot \text{min} (step^{-0.5}, step \cdot warmup^{-1.5}), 
\end{equation}
and the warm-up step is set to $12000$. To avoiding overfitting, we set dropout rate to $0.1$. We use SpecAugment \cite{park2019specaugment} for data augmentation, and we leverage label smoothing with $0.1$ during training. We average parameters of the models which are saved at the last $10$ epochs as the final model. The maximum length $L$ is set to $60$, which is set by counting the characters in the utterances of the training set.

All the systems are implemented with PyTorch \cite{paszke2019pytorch}.All the experiments are conducted on an NVIDIA RTX 2080Ti GPU. For evaluating inference speed, we predict one utterance once for evaluate speed, i.e., the batch contains $1$ utterance. For the autoregressive models, beam-width is set to $5$ and the maximum decoding length is $60$. 

\vspace{-5pt}
\subsection{Results}
\vspace{-5pt}

We first compare the LASO with different numbers of attention blocks of PDS on the development set. \autoref{tab:hyper} shows the results. We can see that different numbers of attention blocks of PDS impact the performance, but the difference is small. In the rest of the experiments, we use $4$ attention blocks in the PDS module, i.e., $N_s=4$. 

\begin{table}[t]
	\caption{ Character Error Rates (CERs) on the development set of the models with different numbers of attention blocks of PDS.   }
	\centering	
\begin{tabular}{|c|c|c|c|c|}
	\hline
	\#block of PDS & 1   & 2   & 3   & 4   \\ \hline\hline
	LASO-base          & 6.4 & 6.5 & 6.5 & 6.4 \\ \hline
	LASO-big          & 6.2 & 6.2 & 6.3 & 6.2 \\ \hline
\end{tabular}
	\label{tab:hyper}
\end{table}

\begin{table}[t]
	\caption{The character error rates (CERs) of the systems on AISHELL-1. Latency is inference time per utterance on test set (including time of feature extraction). Real-time factor (RTF) is computed as the ratio of the total inference time to the total duration of the test set. Inference is done utterance by utterance without batching, on an NVIDIA RTX 2080Ti GPU. }
	\centering	
	\begin{threeparttable}
		\resizebox{\linewidth}{!}{	
			\begin{tabular}{|l|c|c|c|}
				\hline
				\multicolumn{1}{|c|}{\multirow{2}{*}{System}} & \multicolumn{2}{c|}{CER \%} & \multirow{2}{*}{RTF/Latency} \\ \cline{2-3}
				\multicolumn{1}{|c|}{} & Dev. & Test &                  \\ \hline\hline
				KALDI (nnet3) * $\dagger$ $\ddagger$         & -    & 8.6  &  -                \\ \hline
				KALDI (chain) * $\dagger$ $\ddagger$        & -    & 7.4  &  -                \\ \hline
				ESPNet (Transformer) $\dagger$ \cite{karita2019comparative}   & 6.0  & 6.7  &  -                \\ \hline
				A-FMLM    \cite{chen2019non}             & 6.2  & 6.7  &  -              \\ \hline
				Fan et al. \cite{fan2019unsupervised}              & -    & 6.7  &  -                \\ \hline\hline
				Transformer (ours)                & 6.1  & 6.6  &   \hspace{-0.75em} 0.19 /  961ms      \\ \hline
				LASO-base         & 6.4  & 7.3  & 0.0034 / 17ms     \\ \hline
				LASO-base + speed perturb        & 6.0  & 6.8  & 0.0034 / 17ms     \\ \hline
				LASO-big         & 6.2  & 7.0  & 0.0043 / 21ms     \\ \hline
				LASO-big + speed perturb            & \textbf{5.8}  & \textbf{6.4}  & 0.0043 / 21ms     \\ \hline
		\end{tabular}}
		\begin{tablenotes}	
			\footnotesize
			\item[*] from the KALDI official repository.
			\item[$\dagger$] with speed perturbation based data augmentation.
			\item[$\ddagger$] with i-vector based speaker adaptation.
		\end{tablenotes}
	\end{threeparttable}
	\label{tab:perf}
	\vspace{-15pt}
\end{table}

The performances are shown in \autoref{tab:perf}. We can see that the LASO models achieve good performance with very low latency. LASO-base achieves a $7.3\%$ of CER on the test set. LASO-big achieves a $7.0\%$ of CER on the test set. Both LASO-base and LASO-big outperform chain model ($7.4\%$) \cite{povey2016purely}, without speed perturbation. And it is very close to the state-of-the-art transformers ($6.7\%$) and our re-implemented transformer model ($6.6\%$). These results confirm our idea: if the implicit language semantic is captured, prediction of tokens without explicit relationship among tokens is feasible. 

The performance of the bigger model LASO-big is better than LASO-base. The large scale parameters and more layers make the model more powerful to extract token-level semantic representation for each position. To further improve the performance, we augment training data with speed perturbation \cite{ko2015audio}, and retrain the two LASO models. We use factors $0.9$ and $1.1$ to perturb the speed of the audio and combine the augmented data with the original data. With speed perturbation, the CERs of LASO-base and LASO-big are further reduced to $6.8\%$ and $6.4\%$, respectively.

We also show inference speed in \autoref{tab:perf}. We can see that the latency of LASO models is much smaller than autoregressive models. The speed-up is about $50\times$. The non-autoregressive structure makes LASO do not need multi-pass forward computation in beam-search. And the feed-forward structure of LASO makes parallel computation efficient. 

\begin{figure}[!t]\centering
	\includegraphics[width=1.0\columnwidth]{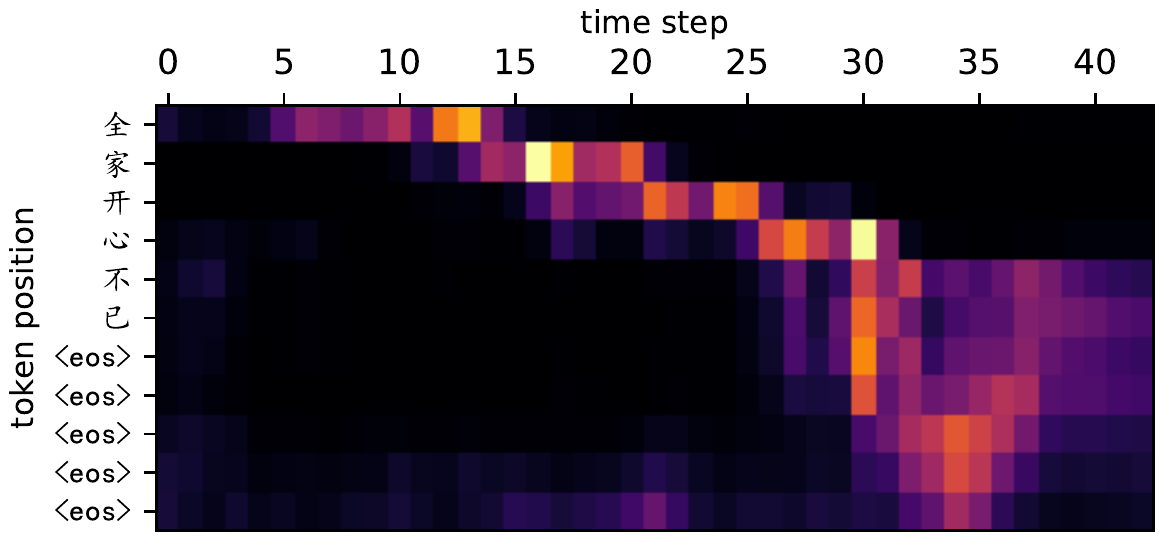} 
	\caption{Visualization of the attention scores of the $4$-th attention block. The horizontal axis is the time step of the outputs of the encoder, and the vertical axis is the token sequence. The token sequence mean ``the family are very happy''. We only show the first $11$ token positions for saving space.} 
	\label{fig:vis}
	\vspace{-15pt}
\end{figure}

To better understand the behaviors of the PDS module, we visualize the attention scores of the $4$-th attention blocks of the PDS in LASO-big. The attention scores are the average of the $8$ heads. We can see that the alignments of the meaningful tokens and the outputs of the encoder are from the upper-left to the bottom-right. For the filler token \texttt{<eos>}, the alignment is vague. Because no certain correspondence between the filler token and the outputs of the encoder exists. Because different head has different alignment in the multi-head attention, the averaged scores are not very sharp.

\vspace{-5pt}
\section{Conclusions and Future Works}
\vspace{-5pt}

\label{sec:conc}
In this paper, we propose a new non-autoregressive speech recognition model. We assume that speech signal contains the relationship among tokens implicitly, and token sequence can be generated without explicit language modeling. Based on this, we propose the LASO model. LASO forward propagates only one-pass for token generation, without beam-search. And because of the feedforward structure, parallel computation can be implemented efficiently, and time cost of inference can be significantly reduced. Experiments demonstrate that the proposed models have very low latency and promising performances. This work is the first result of the LASO model. In the future, we will investigate how to improve the performance of the LASO model by loss functions.

\vspace{-10pt}
\section{Acknowledgements}
\vspace{-5pt}
This work is supported by the National Key Research \& Development Plan of China (No.2016YFB1001404), the National Natural Science Foundation of China (NSFC) (No.61831022, No.61901473, No.61771472, No.61773379) and Inria-CAS Joint Research Project (No.173211KYSB20170061 and No.173211KYSB20190049). This work is also (partially) funded by Huawei Noah's Ark Lab. We also thank the anonymous reviewers for their invaluable comments.

%\newpage
\bibliographystyle{IEEEtran}
\bibliography{mybib}

\end{document}